\documentstyle[aps,twocolumn,psfig]{revtex}

\begin{document}
\title{Superconductivity in a magnetically ordered background}
\author{A. Amici, P. Thalmeier, and P. Fulde}
\address{Max-Planck-Institut f\"ur Physik komplexer Systeme, 
N\"othnitzer Stra{\ss}e 38, 01187 Dresden, Germany}

\maketitle

\begin{abstract}
Borocarbide compounds with the formula RNi$_2$B$_2$C
show interesting superconducting and magnetic properties
and the coexistence of the two phenomena.
BCS theory is extended
to systems with underlying commensurate magnetic order.
In the case of helical phases the
technique may be extended to any ${\bf Q}$-vector 
and there exists a well defined
limit for incommensurate values.
The way magnetic order influences superconductivity
depends crucially on the details of both the magnetic structure
and the electron bands, but some qualitative criteria may
be given. As an example we give a brief analysis of the 
compound HoNi$_2$B$_2$C.
\end{abstract}
\vspace{.5cm}
%\newpage

The borocarbides are a class of compounds with
formula RNi$_2$B$_2$C (R = Y, La, or a rare earth). They attract
attention because of their interesting superconducting and
magnetic properties and their mutual interaction. 
Compounds with Y, Lu, Tm, Er, Ho and Dy 
have phonon mediated superconductivity
at relatively high temperatures \cite{cava94}.
On the magnetic side Gd, Tb, Dy, Ho, Er and Tm show long-range magnetic order
with a variety of different structures \cite{lynn97}.
Crystalline electric field
effects determine the easy direction of the spins in 
the localised $f$-states
of the rare earth ions, while the RKKY interaction among them
determines the ordering ${\bf Q}$ vector.
Commensurate and incommensurate magnetic
structures coexist with superconductivity. The case of HoNi$_2$B$_2$C
has been the most spectacular, where a $c$-axis incommensurate
helix, an $a$-axis incommensurate phase and a $c$-axis antiferromagnet
appear in the same temperature region in which the superconducting
upper critical field shows a pronounced depression \cite{gold94,hill96}.

\begin{figure}
\label{band}
\psfig{file=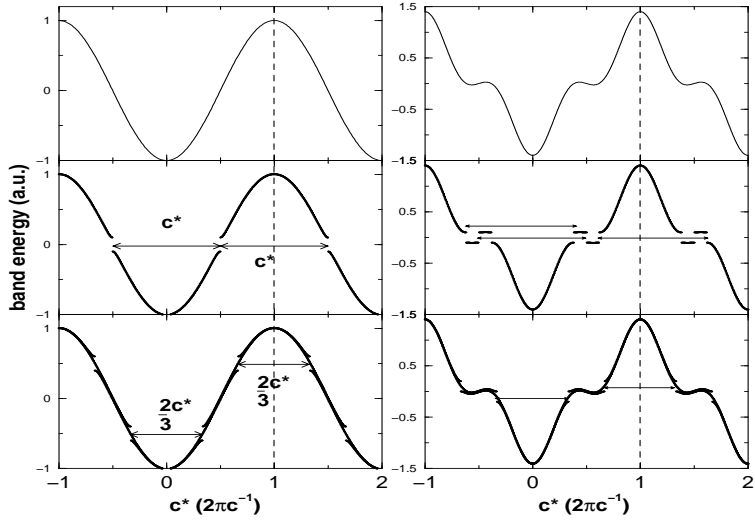,width=10cm,height=7cm}
\caption{Effect of the magnetic order on two model band structures.
The magnetic vector used are $q = \pi$ in the middle
row, corresponding to antiferromagnetic order,
and $q = \frac{2\pi}{3}$ in the lower row, 
corresponding to structure with periodicity
of three unit cells.
Note that magnetic gaps here correspond to $\frac{h}{\epsilon_F}\sim 0.1$
in order to emphasise them, physical values for the gaps are 10 to 100
times smaller.}
\end{figure}

For the description of
superconductivity in the coexistence region we use an extended
BCS theory. The Hamiltonian of band electrons interacting with a given
exchange field is:
\begin{equation}
H_{M} = 
\sum_{{\bf k}\sigma}\epsilon_{\bf k} \, c^{+}_{{\bf k}\sigma}c_{{\bf k}\sigma}-
\sum_{{\bf kq}\sigma\sigma'}{\bf h_q} \cdot
\mbox{\boldmath $\sigma$}^{\sigma\sigma'} \,
c^{+}_{{\bf k+q}\sigma}c_{{\bf k}\sigma'}, 
\end{equation}
where ${\bf h_q}$ is the Fourier transform of the internal effective 
magnetic field (${\bf h_q}=\mu_B\,g\,I({\bf q})\,{\bf S(q)}$) and
{\boldmath $\sigma$}$^{\sigma\sigma'}$ is 
a vector with the Pauli matrices as the components.
Conventional notation is used for the other symbols \cite{allen82}.
This has to be added to the usual phonon Hamiltonian:
\begin{equation}
H_{SC}=\sum_{\bf q}\omega_{\bf q} \, a^{+}_{\bf q}a_{\bf q}+
\sum_{{\bf k'k}\sigma}g_{\bf k'k} \, c^{+}_{{\bf k'}\sigma}c_{{\bf k}\sigma}
\left(
a_{\bf k'-k}+a^+_{\bf k-k'}
\right)
\end{equation}
For a given exchange field ${\bf h_q}$ commensurate with the lattice
the eigenstates problem for $H_M$ is reduced to solving a $n\times n$
matrix, where $n$ is the number of rare earth atoms in the magnetic unit cell.
Using the extended unit cell representation, the total Hamiltonian of the
system is given by:
\begin{eqnarray}
H&=& 
\sum_{{\bf k}\nu}\tilde{\epsilon}_{{\bf k}\nu} \, 
\tilde{c}^{+}_{{\bf k}\nu} \tilde{c}_{{\bf k}\nu}+
\sum_{\bf q}\omega_{\bf q} \, a^{+}_{\bf q}a_{\bf q}+ \nonumber \\
&+&\sum_{{\bf k'k}\nu'\nu}\tilde{g}^{\nu'\nu}_{\bf k'k} \, 
\tilde{c}^{+}_{{\bf k'}\nu'} \tilde{c}_{{\bf k}\nu}
\left(
a_{\bf k'-k}+a^+_{\bf k-k'}
\right)
\end{eqnarray}
The main differences with respect to the non-magnetic case are:
\\ (1) the magnetic eigenstates are labeled by the index $\nu$ 
that does not refer to a definite spin state. Furthermore
energy spin degeneracy is in general lifted and magnetic Cooper pairs
formed with these states will not be in spin singlet state
\\ (2) the band structure acquires gaps at magnetic Bragg planes and
possibly at other characteristic surfaces.
Whenever the Fermi surface intersects
one of those planes or surfaces it is force to become orthogonal to it.
However, apart from pathological situations, the properties of the electrons
are affected only to order $\frac{h}{\epsilon_F}$
\\ (3) the scattering amplitude between magnetic states 
$\tilde{g}^{\nu'\nu}_{\bf k'k}$ acquires further ${\bf k}$
dependence linked to the underlying exchange field.
Typical magnetic energy bands are shown in the figure \ref{band}.

Introducing the gap function matrix:
\begin{equation}
\Delta^{\nu'\nu}=\left<
  \tilde{c}_{{\bf k}\nu'} \tilde{c}_{{\bf -k}\nu}
\right>
\end{equation}
we are able to construct the mean-field theory of
superconductivity in a way quite close to the standard one.
Some qualitative remarks are now possible:
in magnetic states with a net magnetisation (i.e. ferromagnets)
$\nu=\sigma$, spin degeneracy is lifted 
and coexistence is possible only for very small values of the magnetisation.
On the other hand, if the net magnetisation is zero and
the magnetic structure is collinear to a given vector ${\bf \hat{n}}$
(${\bf S(x)} = S({\bf x})\cdot {\bf \hat{n}}$), the spin 
along that direction is conserved. In these cases
(i.e., antiferromagnets \cite{ful81} 
and longitudinally modulated phases) the magnetic
theory reduces to the usual one but with modified band structures
and interaction with phonons.

At a qualitative level it may be noticed that:
\\ (1) if magnetic order does not destroy large pieces of the
Fermi surface, as in the case of nesting, electronic properties
are only slightly modified and the opening of the magnetic gap
does not imply strong suppression of superconductivity
\\ (2) time-reversal symmetry of the system
is not a necessary condition 
for the existence of superconducting pairs, nor is time-reversal 
plus a translation symmetry\cite{ful81}

Now we consider the case of helical magnetic order. For a helix
with wave vector ${\bf Q}$ and amplitude $h$, the 
magnetic eigenstates may be computed analytically
through a Bogoliubov transformation:
\begin{displaymath}
\tilde{\epsilon}_{{\bf k}\pm} =
  \frac{\epsilon_{\bf k\pm Q}+\epsilon_{\bf k}}{2}+
  \frac{\epsilon_{\bf k}-\epsilon_{\bf k\pm Q}}{2}
  \sqrt{1+\frac{4h^2}{(\epsilon_{\bf k}-\epsilon_{\bf k\pm Q})^2}}
\end{displaymath}
and the Bogoliubov coefficients are:
\begin{displaymath}
u({\bf k})=\sqrt{
             \frac{1}{2}\left[
               1+\sqrt{
                 \frac{(\epsilon_{\bf k}-\epsilon_{\bf k+Q})^2}
                      {(\epsilon_{\bf k}-\epsilon_{\bf k+Q})^2 + 4h^2}
               }
             \right]
           }
\end{displaymath}
\begin{displaymath}
v^2({\bf k})=1-u^2({\bf k})
\end{displaymath}
Cooper pairs formed with these electronic states are in a mixed state of
singlet and triplet.
Using the Einstein approximation for the phonon dispersion law ($\omega_0$)
and the structureless electron-phonon interaction ($g_{\bf k'k}=g$) in
the weak coupling limit it is possible to
obtain the standard BCS equation
for the gap function with a modified
definition of the interaction parameter (see Ref. \cite{mor96} for more 
details):
\begin{displaymath}
\tilde{\lambda} = \frac{2g^2}{\omega_0}\int\limits_{MFS} dS \, \frac{\left(
u^2({\bf k})-v^2({\bf k})
\right)^2
}{|\nabla_{\bf k}\tilde{\epsilon}_{\bf k}|}
\end{displaymath}
In order to give a quantitative estimate for the new effective
electron-phonon coupling constant we assume cylindrical 
symmetry around the $z$-axis for the energy bands
and linearise them at the magnetic gap. In this way band effects
enter through the two components of the Fermi velocity ($\hbar{\bf v}_F=
\nabla_{\bf k}\epsilon_{\bf k}$):
the one orthogonal ($v_\perp$) and the one 
parallel ($v_\parallel$) to the $z$-axis.
The difference in the interaction parameter is given by:
\begin{displaymath}
\Delta\lambda = \tilde{\lambda} - \lambda = -\frac{g^2}{\pi\,\hbar^2\omega_0}
\,\frac{k_r}{v_\perp v_\parallel}\,h
\end{displaymath}
where $k_r\sim k_F$ is the radius of the intersection 
between the Fermi surface and the magnetic Bragg plane.
The relative reduction of the superconducting parameter is then:
\begin{equation}
\label{dlambda}
\frac{\Delta\lambda}{\lambda} \sim
- \frac{(\hbar v_F k_r) \, h}{(\hbar v_\perp k_F) \, (\hbar v_\parallel k_F)}
\end{equation}
Eq. (\ref{dlambda}) has the following qualitative features:
\\ (1) assuming $v_\perp$ and $v_\parallel$ of order $v_F$ the variation of
the interaction parameter is proportional to 
$\frac{h}{\epsilon_f}$ and $\lambda$ is reduced only by a few percent
\\ (2) however RKKY interaction maxima are usually connected to
pieces of the Fermi surface orthogonal
to the ${\bf Q}$ vector close to the Bragg planes. 
If the ${\bf Q}$ vector of the magnetic
order is determined  by the RKKY interaction (as it happens generally for 
incommensurate structures), then $v_\perp \ll v_F$ and the 
suppression of superconductivity may be large.
These qualitative features may give an interpretation
of the behaviour of superconductivity in HoNi$_2$B$_2$C. 
From anisotropic magnetisation measurement a broad maximum
in the RKKY function is inferred close to $q=(0,0,\pi)$ \cite{ami98}.
This maximum is related to both the 
magnetic $c$-axis incommensurate helix and to
the $c$-axis antiferromagnet. Because of its flatness it is probably
due to the overall structure of the electronic bands and not to
sharp nesting features, thus $v_\perp\sim v_F$.
On the other hand the $a$-axis modulation appears to be related
to nesting features of the Fermi surface (both experimentally and
through band calculations \cite{rhee95}). We therefore suggest
that the $a$-axis modulation has a
stronger destructive character for superconductivity 
than the $c$-axis modulation \cite{loew97}.

We conclude that an extended BCS theory may be constructed
in the general case of commensurate magnetic structures.
In the special case of helical magnetic order the extended theory
is also valid in the incommensurate limit.
A possible qualitative explanation for the 
anomalies in the superconductivity of HoNi$_2$B$_2$C has been suggested.

\end{document}